\documentstyle[12pt]{article}
\newcommand{\prepr}[1] {\begin{flushright} {\bf #1} \end{flushright}
  \vskip 1.5cm}
\newcommand{\titul}[1] {\begin{center}{\large\bf #1 }
\end{center}\vskip 1.cm}

\newcommand{\abstr}[1] {{\begin{center} \vskip .5cm {\bf Abstract
                        \vspace{0pt}} \end{center}}\begin{quote} #1
                        \end{quote}}

\begin{document}
\begin{titlepage}
\prepr{JINR E2-96-xxx
\\July  1996}
\titul{  QCD analysis  of the CCFR data for  $xF_3$
and Higher--Twist Contribution. }

\begin{center}
{\bf A.V. Sidorov}\\ [1cm]
 {\em Bogoliubov Laboratory of Theoretical Physics\\
 Joint Institute for Nuclear Research, 141980 Dubna, Russia}\\
\end{center}

\abstr{

The QCD analysis of the $xF_3$ structure function measured
in deep-inelastic scattering of neutrinos and antineutrinos
on an iron target at the Fermilab Tevatron is done in
1--, 2-- and 3--loop  order of QCD.
The x dependence of the higher--twist contribution is evaluated.
The experimental value of
higher--twist corrections to the Gross--Llewellyn Smith sum rule
is discussed.
}
\end{titlepage}

At present, the precise measurements of structure functions (SF)
and detailed theoretical calculations of QCD
predictions for scaling violations
( up to 3--loop order for $xF_3(x,Q^2)$ and $F_2(x,Q^2)$)
provide an important means of accurate comparison of QCD with experiment.
The importance of higher--twist (HT)
contribution to SF was pointed from the very beginning
of QCD comparison with experimental data \cite{AbbBar} on SF.
 Despite a fast progress in theoretical
QCD calculations of power corrections to nonsinglet SF and sum rules
\cite{httheor,webber} ( for reviews and references see \cite{RevHT}),
the shape of HT
(order $1/Q^2$) contributions is measured only for $F_2$ SF \cite{htf2}
and is still only estimated for $xF_3$ \cite{Barker}.
In the present note, the x dependence of HT contribution is phenomenologycally
determined in the framework of
QCD analysis of the experimental data of the CCFR collaboration {\footnote{
Announced by CCFR collaboration reevaluation of the structure functions could
change the results of the QCD analysis.}}  obtained
at Fermilab Tevatron \cite{CCFR} for the $xF_3$  structure functions of the
deep-inelastic scattering of neutrinos and antineutrinos on an Iron target
by means of the Jacobi polynomial expansion method in
the 1--, 2-- and 3--loop  order of QCD.

 The details of this method are described
in \cite{PS}-\cite{nnl}. The $Q^2$ - evolution of the moments
$M_3^{QCD}(N,Q^2)$ is
given by perturbative QCD \cite{s4,s5}.

\begin{eqnarray}
M_3^{QCD}(N,Q^2)
& =& \left [ \frac{\alpha _{S}\left ( Q_{0}^{2}\right )}
{\alpha _{S}\left ( Q^{2}\right )}\right ]^{d_{N}}
H_{N}\left (  Q_{0}^{2},Q^{2}\right )
M_3^{QCD}(N,Q^2_0) ,~~~N = 2,3, ...  \label{m3q2} \\
d_N & = & \gamma^{(0),N}/2\beta_0,
. \nonumber
\end{eqnarray}

Here $\alpha_s(Q^2)$ is the constant of strong interaction,
$\gamma^{(0)NS}_{N}$ are nonsinglet leading order anomalous dimensions.
The factor $H_{N}\left (  Q_{0}^{2},Q^{2}\right )$ contains all next--
and next--to--next--to--leading order QCD corrections and is constructed in
accordance with \cite{nnl} based on theoretical results of \cite{thnnl}.

Having at hand the moments (\ref{m3q2})
and following the method \cite{Kri,BCDMS}, we can write
 the structure
function $~xF_3~$ in the form:
\begin{equation}
xF_{3}^{pQCD}(x,Q^2)=x^{\alpha }(1-x)^{\beta}\sum_{n=0}^{N_{max}}
\Theta_n^{\alpha , \beta }
(x)\sum_{j=0}^{n}c_{j}^{(n)}{(\a ,\beta )}
M_{3}^{QCD} \left (j+2, Q^{2}\right ),   \\
\label{e7}
\end{equation}
where $~\Theta^{\alpha \beta}_{n}(x)~$ is a set of Jacobi polynomials and
$~c^{n}_{j}(\alpha,\beta)~$ are coefficients of the series of
$~\Theta^{\alpha,\beta}_{n}(x)~$ in powers of x:
\begin{equation}
\Theta_{n} ^{\a , \beta}(x)=
\sum_{j=0}^{n}c_{j}^{(n)}{(\a ,\beta )}x^j .
\label{e9}
\end{equation}

The unknown coefficients $M_3(N,Q^2_0)$ in (\ref{m3q2}) could be parametrised
as Mellin moments of some function:
\begin{eqnarray}
M_3^{QCD}(N,Q^2_0)&=&\int_{0}^{1}dx{x^{N-2}}Ax^b(1-x)^c(1+\gamma x),
~~~ N = 2,3, ...
\label{Mellf30}
\end{eqnarray}

For $~N_{max} = 8~$ the accuracy better than $~10^{-3}~$ is achieved
in a wide region of parameters $~ \alpha~$ and $~\beta~$~ \cite{Kri}.
In particular, we use $~ \alpha=0.7$ and $~\beta=3.0$~

 Using Mellin moments (\ref{m3q2}),(\ref{Mellf30}), expression (\ref{e7})
for SF and                          taking target--mass corrections (TMC)
into account, we have
reconstructed  $xF_3^{pQCD}(x,Q^2)$. Five free parameters:
A, b, c, $\gamma$ and QCD parameter $\Lambda_{\overline{MS}}$
are to be determine from comparison with experimental data.

To extract the HT, contribution we
parameterize the nonsinglet SF as follows:
\begin{eqnarray}
xF_3(x,Q^2)=xF_3^{pQCD}(x,Q^2)+h(x)/Q^2,
\label{xf3}
\end{eqnarray}
where the $Q^2$ dependence of the first term in the r.h.s is determined by
perturbative QCD. Constants
$h(x_i)$ (one per x--bin) parameterize the HT x dependence. In accordance with
the x-bin structure of the CCFR data we put $x_i=$
$0.015,~0.045,~0.080,$ $~0.125,~0.175,$ $~0.225,~0.275,$ $~0.350,$ $~0.450,
~0.550,$ $~0.650$
for $i=1,2...11$. The values                   of constants $h(x_i)$
as well as parameters A, b, c,$\gamma$  and
scale parameter $\Lambda$       are determined by fitting the
set of the CCFR data at 90 experimental points of $xF_3$ in a wide
kinematical region:
$1.3~GeV^2\leq~Q^2~\leq~501~GeV^2$ and $0.015~\leq~x~\leq~0.65$
 and  $Q^2_0=10~GeV^2$. We have put the number of flavours to equal 4.
The  TMC are taken into account to the order of  $o(M_{nucl}^4/Q^4)$ .
The  nuclear
effect of the relativistic  Fermi motion is estimated
>from below by the ratio $R_F^{D/N}=F_3^D/F_3^N$ \cite{SiTo} obtained
in the covariant
approach in light-cone variables \cite{BrTo}.

Results of the fit are presented in Table 1 and Figures 1-3. The theoretical
prediction for $h(x)$  from \cite{webber} is presented at Figure 3.
\\[3mm]

   \hspace{4mm}

   Several comments are in order:
   \begin{itemize}
\item A decrease of  $\chi^{2 (NNLO)}$ in comparison
 with  $\chi^{2 (NLO)}$ and $\chi^{2 (LO)}$:
$\chi^{2, NNLO}_{d.f.}<\chi^{2, NLO}_{d.f.}<\chi^{2, LO}_{d.f.}$
 demonstrates that 3--loop effects are important for the kinematical
 region under consideration. For all orders of QCD the  $\chi^{2}$
per degree of
freedom is smaller than in \cite{nnl}, where the fit was done without
HT contribution.
\item  The obtained value of the $\Lambda$ is smaller in comparison
with results of the
previous analysis of CCFR data \cite{KaSi,nnl} with the cut off $Q^2>10~GeV^2$
$\Lambda_{\overline{MS}}^{NNL)}=184\pm31~MeV$ but exhibits
relatively large statistical errors. Results of the  NNLO fit gives
the constant of strong  interaction
$\alpha_S^{NNLO} (M_Z^2)=0.104^{+0.006}_{-0.008}(syst.)$
 in agreement within the errors
with usual DIS results \cite{Beth95} and with the predictions of
CCFR-NuTeV Collaboration \cite{alzccfr} based on the test of the
Gross--Llewellyn Smith (GLS) sum rule.
\item
The shape of h(x) demonstrates for LO, NLO and NNLO fit
a very small value at $0.015 \leq x \leq 0.045$, a negative value at
$0.1 \leq x \leq 0.045$  (with a minimum located at about $x=0.2$)
 and increase from
a negative to a positive value at $0.2 \leq x \leq 0.65$. This behavior is in
qualitative agreement with theoretical predictions of \cite{webber}
and reproduces appropriately the predicted zero of
h(x):~ $x^{theor}\sim 0.67$~
while in our NNLO analysis $x^{NNL}\sim 0.40$~.   A separate fit with cuts off
$Q^2>5~GeV^2$   and $Q^2>10~GeV^2$ shows the stability of shape of h(x) and
increase of errors.
\item
The absolute value of h(x) slightly decreases from LO to NNLO fit.
It may be indicates a special role of higher order
perturbative QCD corrections reveals by renormalon technique \cite{BHT}:
at higher order $xF_3^{pQCD}$
in (\ref{xf3}) describes effectively the power corrections.
\item
Definite theoretical predictions are presented for the  first moment of h(x)
which contributes to the
GLS sum rule \cite{gls}:
$
h_1=\int_0^1 \frac{h(x)}{x}dx~.$
A general structure of this contribution is known
>from the results of Ref.\cite{SV}
The corresponding numerical
calculations of this term was made in Ref. \cite{BK}
$h_1=-0.29 \pm 0.14$\footnote{Here by and after we present value of h(x)
in $[GeV^2]$}
and more recently in Ref. \cite{HT}
$h_1=-0.47\pm0.04$,
using the same three-point function QCD--sum--rules
technique.   One can estimate $h_1$ based on the results of Table 1.~:
$h_1^{LO}=0.12\pm0.53$~, $h_1^{NLO}=0.14\pm0.53$
and $h_1^{NNLO}=0.13\pm0.45$. Taking into account the errors
the values of $h_1^{LO}$, $h_1^{NLO}$ and $h_1^{NNLO}$ could be compared
with the prediction of \cite{BK}
and the recent result of \cite{BHT} for GLS sum rule:
\begin{eqnarray}
\mbox{\rm GLS} &=& 3\Bigg\{\Bigg[1 -\frac{\alpha_s(Q)}{\pi} +\ldots
   \pm \frac{0.02 - 0.07\,}{Q^2}\Bigg]
   - \frac{(0.1\pm 0.03)\,}{Q^2}\Bigg\}
   + O(1/Q^4)    \nonumber
\end{eqnarray}

It should be noted that the fit without the nuclear effect
 $R_F^{D/N}=1$ provides
$h_1^{LO,R=1}=0.11\pm0.51$~,
$h_1^{NLO,R=1}=0.12\pm0.40$~ and $h_1^{NNLO,R=1}=0.12\pm0.48$~
in a good  agreement   with previous results.
The large contribution of small x region to $h_1$
needs the shadowing correction
taking into account for more detail analysis \cite{FS}.
\end{itemize}

In conclusion it should be stressed, that for precise determination of the HT
contribution to SF the role of nuclear effect should be clarified and
a more realistic approximation for $R_F^{Fe/N}=F_3^{Fe}/F_3^N$
is needed. A possible interplay of the nuclear effect and TMC
was considered in \cite{TMCHT}.
We also did not take into account the
threshold effects on $Q^2$ evolution of SF due to
heavy quarks \cite{match} which is necessary owing to a wide
kinematical region
of data under consideration. \\ [4mm]

{\bf Acknowledgements.} \\[2mm]

The author is grateful to Prof. M.V.~Tokarev
for discussions.
This investigation has been
supported in part by INTAS grant No 93-1180
and by the Russian Foundation for Fundamental
Research (RFFR) N 95-02-04314a.

\newpage

Figure captions. \\[5mm]

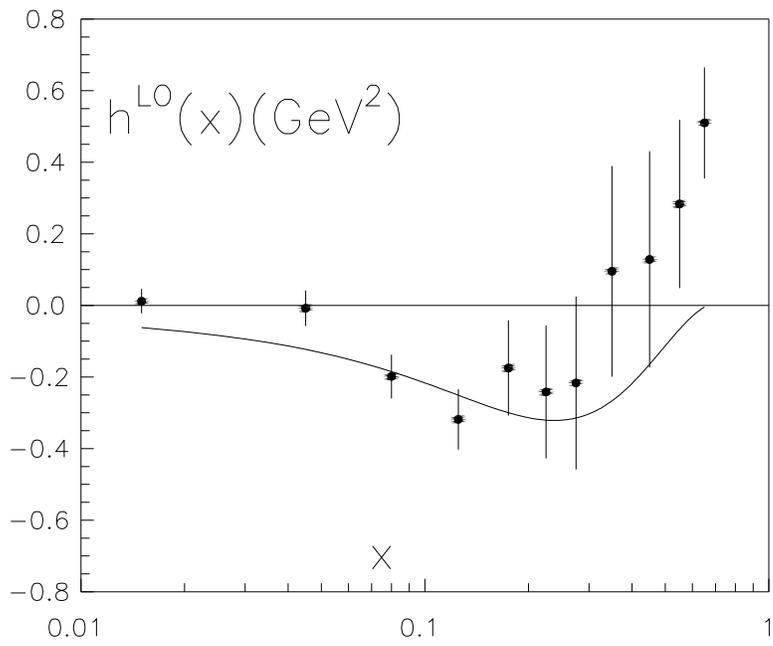
Fig.1.  Higher--twist contributions from LO fit and the
theoretical prediction for $h(x)$ from \cite{webber}. \\[5mm]

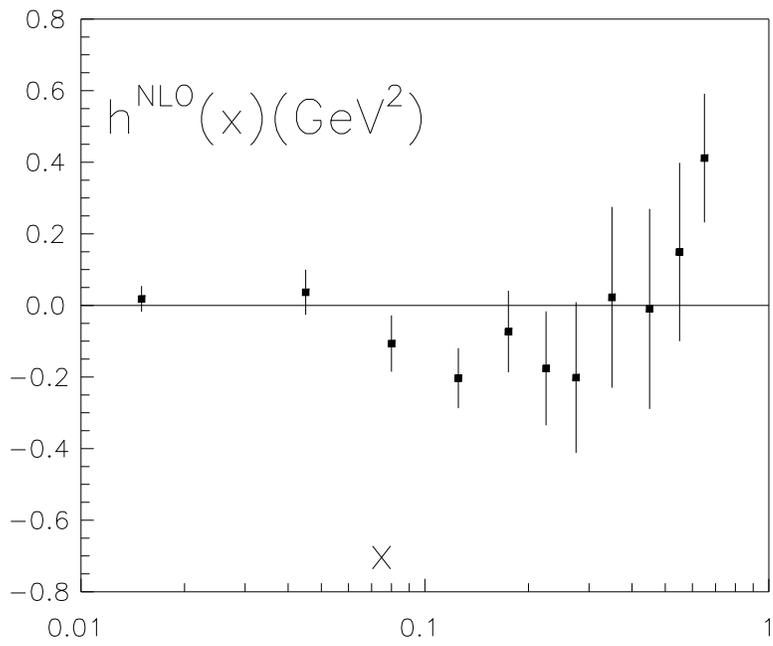
Fig.2.  Higher--twist contributions from NLO fit.  \\[5mm]

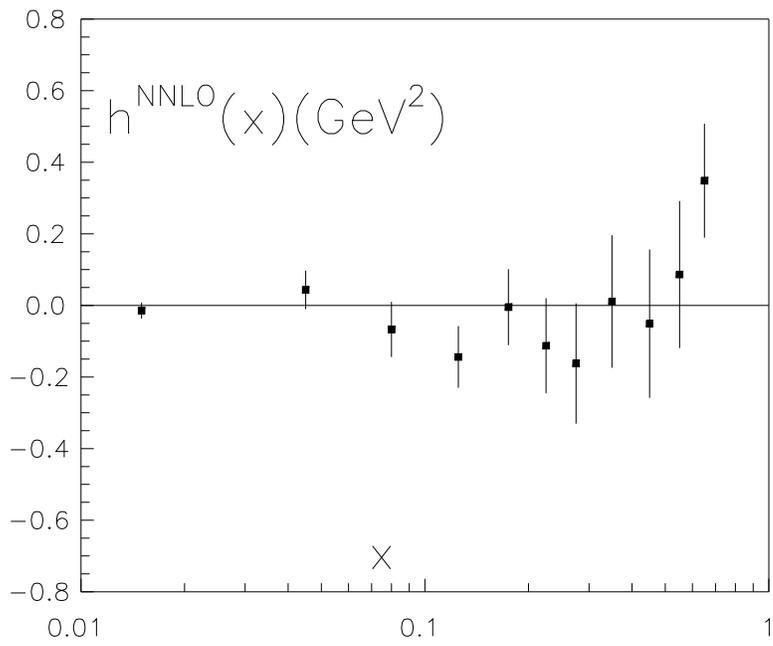
Fig.3.  Higher--twist contributions from NNLO fit. \\[5mm]

\newpage
Table I. Results of 1-, 2- and 3- order
QCD fit (with TMC)
of the CCFR $xF_3$ SF data for  $f=4$, $Q^2>1.3 GeV^2$ with
the corresponding statistical errors
and values of $h(x)$ at different values of x.
$N_{MAX}=10$  for 1- and 2- oder and $N_{MAX}=7$ for 3- order fit.
\newpage

Table I.

\begin{center}
\begin{tabular}{||c|c|c|c||} \hline
                                &      LO              &
    NLO          &         NNLO               \\ \hline
     $\chi^2_{d.f.}$              &   65.1/74                 &
    62.9/74               &     60.9/74            \\
        A                         &    6.69  $\pm$     0.87   &
    6.04  $\pm$     0.51    &      5.56       $\pm$    0.18      \\
        b                         &   0.772  $\pm$     0.040  &
    0.745  $\pm$     0.026   &     0.719       $\pm$    0.011     \\
        c                         &    4.04  $\pm$     0.16   &
    3.97  $\pm$     0.14    &      3.91       $\pm$    0.12      \\
    $\gamma$                      &   0.424  $\pm$     0.53   &
    0.603  $\pm$     0.317   &     0.707       $\pm$    0.055     \\
$\Lambda_{\overline{MS}}$         &    76    $\pm$     62     &
    132    $\pm$     80      &       134       $\pm$     57          \\
                         $[MeV]$  &                           &
                            &     \\ \hline \hline
        $x_i$
 &\multicolumn{3}{c||}{  $h(x_i)~[GeV^2]$ }                                                     \\  \hline
0.015    &   0.012  $\pm$     0.034  &     0.018  $\pm$     0.036   &
   -0.015  $\pm$     0.022      \\
0.045    &  -0.008  $\pm$     0.049  &     0.037  $\pm$     0.063   &
    0.043  $\pm$     0.054      \\
0.080    &  -0.199  $\pm$     0.061  &    -0.107  $\pm$     0.079   &
   -0.067  $\pm$     0.077      \\
0.125    &  -0.318  $\pm$     0.084  &    -0.203  $\pm$     0.083   &
   -0.144  $\pm$     0.086      \\
0.175    &  -0.175  $\pm$     0.133  &    -0.073  $\pm$     0.114   &
   -0.005  $\pm$     0.106      \\
0.225    &  -0.242  $\pm$     0.186  &    -0.176  $\pm$     0.159   &
  -0.113  $\pm$     0.133      \\
0.275    &  -0.217  $\pm$     0.241  &    -0.202  $\pm$     0.210   &
  -0.162  $\pm$     0.168      \\
0.350    &   0.095  $\pm$     0.294  &     0.023  $\pm$     0.253   &
  0.011  $\pm$     0.185      \\
0.450    &   0.129  $\pm$     0.302  &    -0.010  $\pm$     0.280   &
   -0.051  $\pm$     0.207      \\
0.550    &   0.283  $\pm$     0.235  &     0.150  $\pm$     0.249   &
    0.086  $\pm$     0.205      \\
0.650    &   0.510  $\pm$     0.155  &     0.412  $\pm$     0.180   &
   0.349  $\pm$     0.159      \\ \hline
\end{tabular}
\end{center}
\hspace{4mm}
\end{document}